\documentclass{aa}
\usepackage{graphicx}
\usepackage{txfonts}
\usepackage{natbib}
\begin{document}
\title{Spectrum of the unresolved  cosmic X ray background: what is unresolved 50 years after its discovery.}
\author{
A. Moretti\inst{1}, S. Vattakunnel\inst{2}, P.Tozzi\inst{2}, R. Salvaterra\inst{3}, P. Severgnini\inst{1}, D. Fugazza\inst{1}, F. Haardt\inst{4,5}, R. Gilli\inst{6}} 

\offprints{alberto.moretti@brera.inaf.it}

\institute{
INAF, Osservatorio Astronomico di Brera, Via Brera 28, 20121, Milano, Italy
\and
INAF, Osservatorio Astronomico di Trieste, Via Tiepolo 11, 34143 Trieste, Italy
\and
INAF, Istituto di Astrofisica Spaziale e Fisica Cosmica,  Via E. Bassini 15, 20133 Milano, Italy
\and
 DiSAT, Universit\`a dellÕInsubria, via Valleggio 11, 22100 Como, Italy
 \and
 INFN, Sezione di Milano Bicocca, P.za Della Scienza 3, 20126 Milano
 \and
 INAF, Osservatorio Astronomico di Brera, Via Bologna, via Ranzani 1, 40127, Bologna, Italy
%
}
\date{Received ; accepted }
\date{Received ; accepted }
\titlerunning{ }
\authorrunning{Moretti et al.}  
\abstract{} 
{We study the spectral properties of the unresolved cosmic X-ray background (CXRB) in the 1.5-7.0 keV energy band 
with the aim of providing an observational constraint on the statistical 
properties of those sources that are too faint to be individually probed.}
{We made use of the Swift X-ray observation of the Chandra Deep Field South complemented by the Chandra data.
Exploiting the lowest instrument background  (Swift)  together with the deepest observation ever performed (Chandra) 
we measured the unresolved emission at the deepest level and with the best accuracy available today.}
{We find that the unresolved CXRB emission can be modeled  by a single power law with a very hard photon index $\Gamma$=0.1$\pm$0.7 
and a flux of  5$^{+3.2}_{-2.6} \times$10$^{-12}$ erg s$^{-1}$ cm$^{-2}$ deg$^{-2}$  in the 2.0-10 keV energy band (1$\sigma$ error).  
Thanks to the low instrument background of the Swift-XRT, we significantly improved the accuracy with respect to previous measurements.}
{These results point towards a novel ingredient in AGN  population synthesis models, namely  
a positive evolution of the Compton-thick AGN population from local Universe to high redshift.}


\authorrunning{A.\ Moretti et al.}

\titlerunning{CXRB: what's left unresolved 50 years after the discovery.}

\maketitle

\section{Introduction}
\noindent
The flux limit reached by the Chandra deep field  South (CDF-S) 4 Ms observation is such that the observed source angular density is four
sources per square arcminute \citep{Lehmer12}.
These sources  have been found to be mostly unobscured and Compton-thin (N$_{\rm H}<$ 10$^{23}$ cm $^{-2}$) AGN at redshift 
$\lesssim$ 2 with some contribution at soft energies ($<$ 2 keV) from galaxy clusters and starburst galaxies 
\citep{Brandt05,Tozzi06,Luo08,Xue11}.
Still, it is worth investigating whether a non-negligible contribution to the
cosmic X-ray background (CXRB) has not been identified yet. This may be due to an
intrinsically diffuse component or to the cumulative contribution of
individual sources below the current flux limit. 
In the latter case, high redshift and Compton-thick AGN
(the ones for which the neutral hydrogen column density is higher than the inverse of the Thomson cross-section,
N$_{\rm H}>$ 1.5$\times$10$^{24}$),  together with star forming galaxies,  are expected to be the main contributors.

\noindent  The very fact that bright quasars, powered by supermassive black holes,  have 
been found at z$>$6  \citep{Fan06,Willott10, Mortlock11}, implies the existence of a
large numbers of  less massive AGN at  these early epochs \citep{Volonteri10}.
Thus, because  no source  at z $>$ 6 has  been detected in the CDF-S, the unresolved CXRB  can be used 
to probe SMBH formation models at these early epochs  \citep{Salvaterra12}. 
In the same way, the unresolved CXRB can be used to put  some constraints  on the contribution 
of the X-ray sources to  re-ionization  \citep{Dijkstra04, Salvaterra05, Salvaterra07, Mcquinn12}. 

\noindent Although  CDF-S observation reaches a flux limit  such that  starburst galaxies  match  
AGN in surface density \citep{Lehmer12}, most of the galaxies still remain undetected  at low redshift, 
contributing to the unresolved CXRB.
Indeed, using the X-ray galaxy luminosity as proxy \citep{Mineo12b},
CXRB can be used to study the evolution of the star formation rate up to redshift  z=6 
\citep{Dijkstra12,Cowie12}.

\noindent  Compton-thick (CT)  are expected to be the main contributors to the unresolved X-ray emission
in the CDF-S. In fact, while their number is expected to be similar to moderately obscured AGN  \citep{Gilli07,Treister09}, 
only a few have been found among the 740 sources of the CDF-S \citep{Tozzi06, Luo11}

\noindent From an observational point of view two different approaches have been pursued to probe the unresolved emission in the 
CDF. The first approach consists in calculating the difference between the total CXRB and  the integrated flux 
of all detected sources. \cite{Moretti03} and \cite{Worsley06}, following this approach, found that the unresolved fraction was  
10-20\%, and higher at harder energies. Moreover, \cite{Worsley04} recognized the signature of a highly absorbed AGN in the spectral 
shape of the unresolved emission.
Unlike these, \cite{Luo11}, while  measuring similar values in the softer band,  found that the unresolved 
emission is consistent with zero, in the 6-8 keV band, posing a strict upper limit. 
The main sources of uncertainty here are the total CXRB measure  \citep{Revnivtsev05, Frontera07, Moretti09} and the statistical 
contribution of the rare and bright sources.

\noindent The second approach is a direct measure of the unresolved emission: \cite{Hickox07}
found that the flux, with large uncertainties ($\sim$200\%),  was consistent with zero in the 2-5 keV band  
both in the CDF-S and CDF-N.
The source of this huge uncertainty is the very low surface brightness of the unresolved emission, which represents only 
a small fraction of the Chandra instrument background. 
In this regime the signal/background ratio is so low that a few percentage points of systematic error in the background measure irreparably
affect the measurement accuracy. 
\footnote{This is exactly the same situation as was faced in the study of the external regions of 
galaxy clusters where the intergalactic medium (ICM) surface brightness is at the same level. In this kind 
of study telescopes such as Suzaku and Swift are preferred to Chandra and XMM, which have much higher instrument background 
and a less favorable ratio signal/background  \citep{Bautz09,Ettori10} .}

\noindent In this paper we work out a different approach, directly measuring the unresolved emission on the Swift XRT 
observation of the CDF-S and subtracting the signal of the sources revealed by Chandra.
In this way we exploit, at the same time, the  low and predictable 
instrument background of the Swift X-ray telescope (XRT) and the unprecedented depth of the Chandra  observation. 

\noindent The paper is organized as follows.
In Section~\ref{sect:data_red} we briefly describe the Swift XRT telescope and the datasets we used, showing their distinctive qualities.
In particular we dwell on the instrument background comparing Swift XRT and Chandra in order to emphasize the peculiarity
of our work.  In Section~\ref{sect:data_ana} we deal with a wealth of technical details to give the full particulars of the spectral analysis 
used in this paper.  For the sake of clarity the fit procedure  and the results are presented in Sect.~\ref{sect:procures}. 
Finally, in Section~\ref{sect:disc} we discuss our results, first, comparing them with previous measurements and 
then, with the expectations from one AGN population synthesis model.

\noindent Throughout this paper, errors are quoted at the 68\% confidence level for the
two parameters of interest ($\Delta\chi^2$=2.3), unless otherwise specified.
We adopt the following cosmology parameter values:
$\Omega_{\rm m} = 0.3$, $\Omega_\Lambda = 0.7$, $h_0 = 0.7$.

\section{Instrument, data, and reduction procedures}\label{sect:data_red}
%
\subsection{Swift-XRT}
The X-ray telescope (XRT) on board the Swift satellite \citep{Gehrels04}, uses a Wolter I mirror set, originally designed for
the JET-X telescope \citep{Citterio94}, to focus X-rays (0.2-10 keV) onto a XMM-Newton/EPIC MOS CCD detector \citep{Burrows05}.  
The effective area of the telescope ($\sim$ 150 cm$^2$ at 1.5 keV) is about five times smaller than Chandra.
The PSF, similar to XMM-Newton, is characterized by a half energy width (HEW) of $\sim$ 18\arcsec  at 1.5 keV \citep{Moretti07}.
The nominal field of view has a radius of $\sim$ 11 arc minutes, with a pixel scale of 2.37\arcsec
\begin{figure}
\includegraphics[width=9cm] {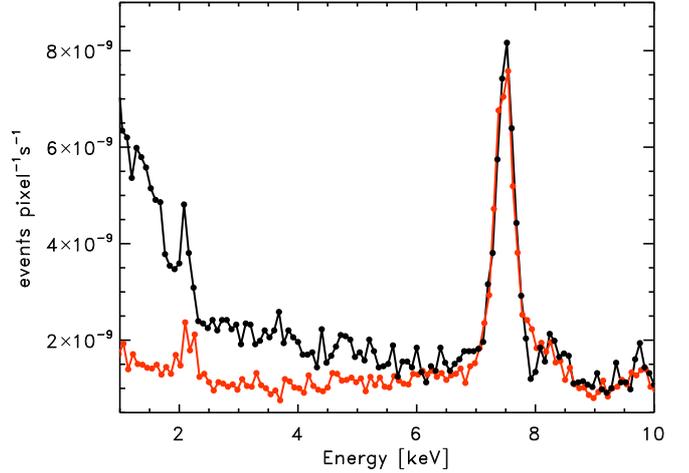}      
\caption{The energy channel (PI) distribution of the XRT unresolved emission (black) compared with the instrument background (red). 
In this plot PI channels have been transformed in energy using a single value and not the RMF matrix. 
 }
\label{fig:nxb}
\end{figure}
%
\subsection{XRT observation of the CDFS}
Swift observed the CDFS for a total nominal exposure of 563700 seconds in the period July 2007 to December 2007. 
We reduced data using the standard software (HEADAS software, v6.10, CALDB version Sep10) and basically following the procedures described 
in the instrument user guide \footnote{http://heasarc.nasa.gov/docs/swift/analysis/documentation} with some minor  modifications for our 
purposes. 
In our analysis we excluded the external (detx $<$ 90 and detx $>$510 )
CCD columns that are affected by the presence of out-of-time-events from corner calibration sources. 
This corresponds to a nominal field of view (FOV) of 16.5 $\times$ 18.9 arcmin (0.087 deg$^2$). 
To detect sources we ran the task {\it wavedetect} \citep{Freeman02} with default input parameter values 
on three different energy bands (0.5 -2.0; 2.0-7.0; 0.5-7.0 keV), on a large number of spatial scales (2,4,8,12,16,24,32,64,128 pixels,
1 pixel equals 2.36\arcsec). 
Merging the three catalogs yields a total of 109 sources down to a limit of $\sim$ 3 $\times$10$^{-16}$ erg s$^{-1}$ cm$^{-2}$ 
and $\sim$ 1$\times$10$^{-15}$ erg s$^{-1}$ cm$^{-2}$ in the 0.5-2.0 and 2.-10. keV bands respectively.
%
\subsection{XRT instrument background }
\label{sect:instr_bkg}
\begin{figure}
\includegraphics[width=9cm] {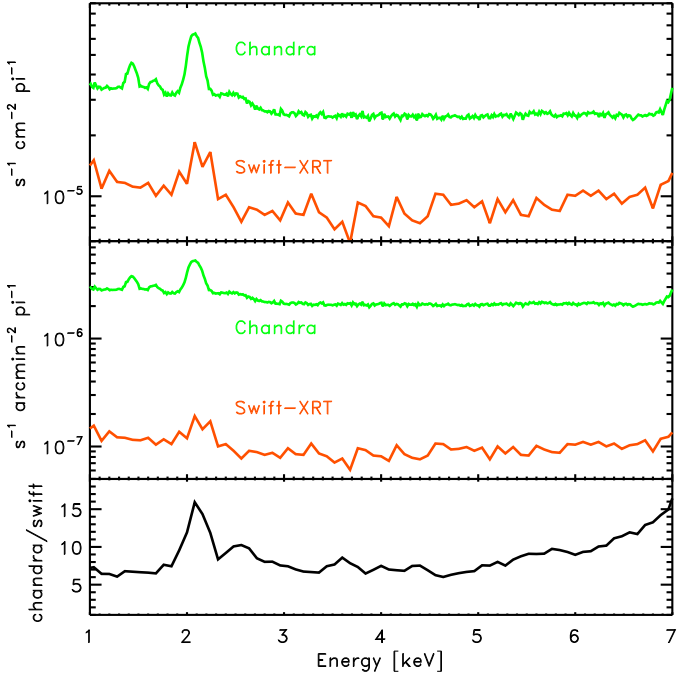}      
\caption{Comparison between Chandra and Swift-XRT instrument background spectra. {\bf Upper panel:} NXB of the two instruments
are plotted per unit area, to show the influence of different environments (orbits plus satellite).  
{\bf Middle panel:} NXB are plotted per solid angle, to show the influence of different focal lengths.  
{\bf Lower panel:} the ratio between Chandra and XRT  NXB calculated  per solid angle and normalized by the respective effective areas.
This shows the improvement in the signal/background ratio for an extended  source using XRT instead of Chandra.}
\label{fig:nxb_comp}
\end{figure}
\noindent For our purposes, the most relevant characteristic of the XRT is the low level and high reproducibility of the instrument background (NXB),
which we define as the signal registered on the CCD when focused radiation is excluded. This is mostly 
contributed by the induced particle emission and the pure instrument electronic noise \citep{Hall08}. 

\noindent To estimate the NXB spectrum in the CDF-S observation, we used  
the signal registered on the small regions in the four CCD  corners that are not 
exposed to the sky (NES regions) and not contaminated by calibration
sources. As these data are only available for observations after June 2008, 
 we used  observations performed in 2008-2009 for a total of $\sim$ 7 Msec. 
For a detailed study of the XRT NXB also see \cite{Moretti09, Moretti11}.
To check at what level of accuracy we can reproduce the XRT instrument background we compared
 this spectrum with the XRT CDF-S observation in the 7-10 keV energy band.
There the effective area is a factor five to ten times lower  than the one at energies $<$2keV and, 
at the same time, the instrument 
background is inflated by the presence 
of the Ni (K$\alpha$ and K$\beta$ at 7.478 and 8.265 KeV) and Au (L$\alpha$ at 9.713 keV) fluorescence lines.
Once the detectable sources have been removed, the cosmic signal,  is expected to be only $\lesssim$ 1\%
 of the background in this  particular energy band.

\noindent As shown in Fig.\ref{fig:nxb} we find that the two spectra coincide very well beyond 7 keV.
Indeed, the ratio of the signals between 7-10 keV is XRT-CDF-S/NES = 1.033, while if we restrict to 7 and 8 keV, 
where the nickel K lines are completely overwhelming, the agreement is even better XRT-CDF-S/NES = 0.993. 
As such,  in the systematic error calculation we conservatively  assumed that the 
instrument background is known at 5\% accuracy (1$\sigma$).
We note that, in some previous works that used Chandra and XMM data  \citep{Deluca04,Hickox06},  the ratio of the signals at the very end of the 
energy band has been used to rescale the measurement of the instrument background. Here we find very good agreement between 
the two datasets, without any ad--hoc renormalization. This plainly shows the high level of reproducibility of the XRT instrument background.   
\subsection{Comparison with Chandra }
\noindent To compare the Swift-XRT and Chandra instrument backgrounds,  we considered the signal registered on the
 XRT NES regions and the signal registered on the ACIS-I in stowed position\footnote{File acis-D-012367-stowed-psu-evt-041104.fits downloaded from
http://cxc.harvard.edu/contrib/maxim/acisbg/}.
 In the upper panel of Fig.~\ref{fig:nxb_comp} we plot the two instrument background spectra normalized 
per unit of area, which show the effect of the different environments (orbit plus satellite). The very fact of being 
in high orbit (assuming negligible effects from the satellite structure differences) produces a particle background 
in Chandra that is a factor $\sim$5 higher than in XRT, which is in a low orbit. 
In the spectroscopic study of extended sources, what is relevant is the background per unit of solid 
angle (middle panel of Fig.~\ref{fig:nxb_comp}), which can be calculated  from the previous one, by accounting 
for the pixel scale. Since the ratio of the pixel linear scale between  Chandra and Swift is $\sim$ 4.5, we find
that the Chandra background per unit of solid angle is $\sim$ 50 higher than Swift XRT. 
Finally, if we take the effective area into account  (which is approximately a factor 5 higher in Chandra), we find
that the ratio signal/background for extended sources (for which the angular resolution is not important)  
is a factor $\sim$10 better in Swift-XRT than in Chandra (lower panel of Fig.~\ref{fig:nxb_comp}).

\section{Spectral analysis}\label{sect:data_ana} 
We restricted our analysis to the sky circular region centered on ra=53.1092 and dec=-27.82086 with  a radius of 0.1053 degrees (0.03184 deg$^2$),
which optimizes the intersection of the two instrument exposure maps. Hereafter we refer to this region as the region of interest (ROI) region.
The average effective exposures within this region is  508  ksec .
Moreover, we restricted our analysis to the 1.5-7.0 keV energy band.
We excluded the soft part ($<$1.5 keV)  because in this range 
our data did not allow us to disentangle the genuine extragalactic CXRB components 
from the local ones. In fact, in this energy band, the diffuse X-ray emission  is expected to be contributed both by the  
thermal emission from the Galaxy and the local Bubble \citep{Snowden98, Kuntz00} and by the thermal 
emission from faint galaxy groups and WHIM filaments \citep{Cappelluti12,Shull12}.
Above 7 keV, as said in the previous section, the instrument background and, in particular, 
the particle-induced one overwhelms the cosmic signal.

\noindent
To study the spectrum of the unresolved CXRB on the Swift-XRT data, first, we excised  the signal of all the detected sources, 
which are 37 within the ROI, with  fluxes in the range [1e-15,1e-14] erg s$^{-1}$ cm$^{-2}$ in the 2.0-10.0 keV band. 
To get rid of most of the source signal without losing 
too much area, we excluded a circular region with a radius that depends on source counts. 
We used a radius equal to twice the radius at which the source surface brightness profile matches the background: this 
means a typical radius of  25 arcsec with values ranging from 20 to 60. Hereafter we refer to this  radius as r$_{ext}$ .

\noindent
Besides the instrument background, the unresolved signal on the XRT detector in the 1.5-7.0 energy band is expected to be  contributed by the following elements:  
\newline -- the sources detected by Chandra and not by XRT,
\newline -- the optically/IR detected (X-ray undetected) sources,
\newline -- PSF residuals from detected sources,
\newline -- the stray-light contamination, 
\newline -- the unresolved CXRB that is the goal of the present work.  

\noindent
To perform spectral analysis we subtracted the instrument background and modeled the remaining five components.
In the following we describe item by item the way we quantified  each single component.
%

\subsection{Chandra sources}

\noindent  We used the public catalog of \cite{Xue11},  which is the result of the cross-correlation of three 
different energy bands and which consists of 740 sources down to the unprecedented flux limits of 
10$^{-17}$ erg s$^{-1}$ cm$^{-2}$  and 10$^{-16}$ erg s$^{-1}$ cm$^{-2}$ 
in the 0.5-2.0 and 2.0-10.0 keV band.
In the selected region the Chandra catalog lists 326 sources that have not been detected by XRT.
We summed the \texttt{pha} files of the sources and of the 
(locally extracted) backgrounds, weighting the response (RMF) and ancillary (ARF) files by the source counts.  
The full details of the reduction of the X-ray spectral data are described in \cite{Vattakunnel12}.

\noindent To quantify the  impact of cross-calibration between Chandra and Swift XRT and source variability 
(Chandra and Swift XRT observation were performed in different periods), we compared the summed spectra of 
the sources detected by both instruments, excluding the five brightest. 
These are 32 sources with  fluxes ranging  between  10$^{-16}$ and 10$^{-14}$ erg s$^{-1}$ cm$^{-2}$  in the [0.5-2.0] band.
The spectral analysis of the two stacked spectra yielded  results that are consistent within 1$\sigma$ at the level of 7\% (Tab.\ref{tab:cross}).
This variance can be easily encompassed in the systematic error budget of the analysis.
This should be considered as an upper limit as the total flux variation associated to the intrinsic variability of the 294 remaining fainter source 
sample is expected to be significantly lower than the 32 brighter ones.    
At the end we conservatively adopted a 7\% (1$\sigma$) value in the systematic error uncertainties in the cross-calibration between the two telescopes.
\begin{table}
\caption[]{Spectral analysis results of 32 stacked XRT sources and corresponding Chandra sources.}
\begin{tabular}{|l|cc|}
\hline
                  & slope                             & flux1-2 keV[erg s$^{-1}$ cm$^{-2}$]\\ 
\hline

Chandra      & 1.09$\pm$(0.01,0.01)    &  2.59$\pm$(0.03,0.03)E-14 \\
XRT            & 1.16$\pm$(0.07,0.08)    &  2.71$\pm$(0.19,0.18)E-14 \\
\hline
\end{tabular}
\label{tab:cross} 
\end{table}
%
\subsection{Optical/IR  galaxies}
\begin{figure}
\includegraphics[width=9cm] {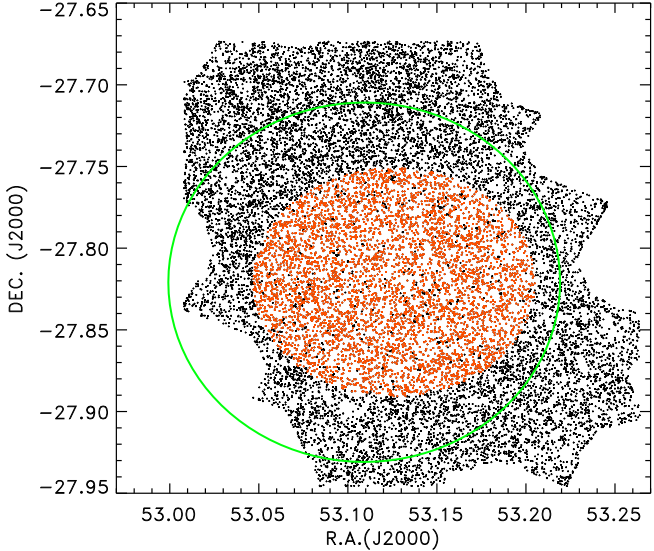}      
\caption{Black dots are the 18296 sources from \cite{Grazian06} catalog. The green circle indicates our ROI.  
Red dots are the 5683 optical/IR sources we used in the spectral analysis.}
\label{fig:optfov}
\end{figure}
\noindent Several studies put in evidence that optical/IR galaxies, which are not detected in X-ray,   
show significant emission once their contribution is averaged (or stacked) 
over a large sample \citep{Worsley06, Hickox07}.
We used  the \cite{Grazian06} multi-wavelength (from 0.3 to 8.0 $\mu$m) catalog, 
which covers the large and deep area in the GOODS Southern Field by combining public data from Hubble  VLT, Spitzer, and  2.2ESO. 
We used the 5683 galaxies detected within the four arcminute radius circular region centered on R.A.=53.126 and  Dec=-27.820, offset by
1 arcminute  to the west from the center of the ROI (Fig.~\ref{fig:optfov}). This ensures a coverage of 40\% of our ROI. 
We found that these galaxies make a low, but significant contribution to the X ray emission, with a ratio signal/background of $\sim$3\%.
To test the systematic uncertainty of this operation we performed the same analysis by varying the region (from 3 to 6 arc minutes of radius) 
and the extraction radius for the single sources (from 2 to 4 pixels). We found that scatter of the result is 10\%  at 1$\sigma$ level, which 
is much less than the statistical error. 
\subsection{PSF residuals}
\begin{figure}
\includegraphics[width=8cm] {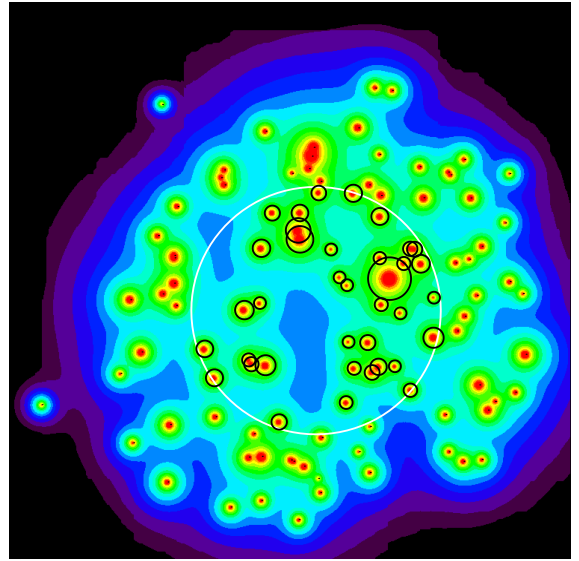}      
\caption{The PSF model of the 109 sources detected in the XRT observation at 2.5 keV energy. The large white circle shows the 
boundaries of our ROI, while small black circles show regions we excised (r$\rm_{ext}$). The PSF residuals is the small, but not
negligible signal outside small circles and within the ROI. }
\label{fig:psfres}
\end{figure}
As said,  from XRT data we excluded most of the signal of the detected sources by excising circular regions, with the size depending on the source 
flux (r$_{\rm ext}$). This leaves a small fraction ($\sim$ 5\%) of source fluence spread on the rest of the field of view, which we evaluated 
using the PSF analytical model \citep{Moretti05}.  We also considered the small contribution from PSF wings from sources 
detected outside the ROI (76 objects). This fraction depends on the energy since harder photons are more scattered. 
We calculated the residuals on a grid of 50 different energies in the 2.0-7.0 keV range. 
We found that, at 2.5 (6.0) keV, 3.7\% (4.0\%)  of the 109 source flux is diffuse all over the ROI (Fig.~\ref{fig:psfres}).
To account for this energy dependence we built an ad hoc ARF file, linearly interpolating the grid test energies,  
such that fitting  the stacked spectrum of the 109 sources  yields the appropriate corrections. 
In the systematic error calculation we assumed that the PSF model extrapolated  up to $\sim$6 arcmin radii is accurate at the level of   
10\% accuracy (1$\sigma$), \citep{Moretti05}.
\subsection{Stray-light contamination}
One of the main components in the unresolved signal in the Swift-XRT data is the stray-light contamination.
This is produced by photons coming from sources that are outside the telescope FOV at
distances between 25 and 100 arcminutes from the optical axis of the telescope.  
A fraction of the photons produced by these sources reach the detector after only one reflection on the
mirror or even directly, passing through the mirror shells without any
interaction.  Some X-ray telescopes mount baffles on top of the mirrors to prevent this contamination.  
This is not the case for XRT, for which the stray light is a significant fraction of the diffuse radiation registered on the CCD.
\begin{figure}
\includegraphics[width=8cm] {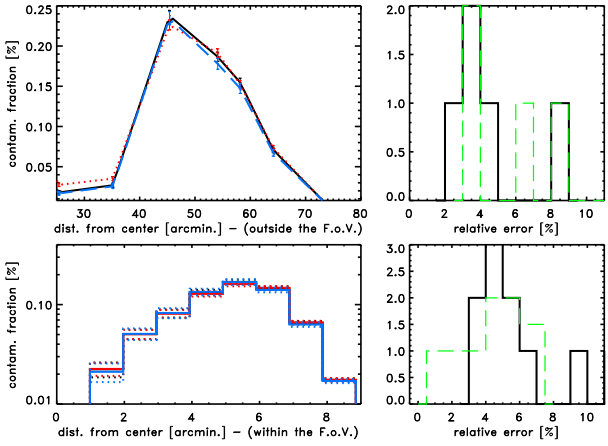}      
\caption{{\bf Upper-left panel:} Fraction of the flux of a source outside of the field of view falling within the annulus between 5 and 6 arc minutes. 
Black, red, and blue points represent the fraction measured in the Crab calibration observations in the total (1.5-7.0), soft (1.0-2.0), and hard bands
 respectively. Lines are the linear interpolation between the measures.
{\bf Upper-right panel:} Black line is the calibration measure error histogram. Green dashed line is the histogram of the relative
difference, between soft and hard band.
{\bf Lower-left panel:} The stray-light contamination on the XRT detector from a uniform source in different concentric annuli on the detector, 
accounting for the CDFS exposure map.  {\bf Right-lower panel: } Black line is the error histogram. 
Green dashed line is again  the histogram of the soft/hard relative difference.} 
\label{fig:slcom}
\end{figure}
\noindent This contamination has been accurately calibrated in \cite{Moretti09} using a set of observations of bright sources (Crab, Sco-X1) 
observed outside the field of view. 

\noindent  To calculate the stray-light contamination in the XRT CDF-S observation, we first  split the detector into 12 concentric annuli 
(external - internal radius = 1\arcmin)  and calculated which fraction of the flux of the calibration source, positioned at a given 
angle outside the field of view, falls within the ith annulus (hereafter fraz$_i$($\theta$)).
For example, we found that, when the Crab is positioned at  45\arcmin~ from the telescope aim,
(0.25$\pm$0.01)\% of  its flux  falls within  the  5-6\arcmin annulus (upper-left  panel of Fig.~\ref{fig:slcom}).
The typical 1 $\sigma$ relative error is $\sim$5\% (upper-right  panel of Fig.~\ref{fig:slcom}).
 As shown in the same figure we do not detect any dependence of the stray-light fraction on energy:  the relative differences between the soft 
 (1.0-2.0 keV) and hard bands (2.0-7.0) keV, defined as abs(fraz$_{soft}$-fraz$_{hard}$)/fraz$_{soft} $, are at the level of the statistical error ($\sim$5\%). 
 
\noindent  Then, for each annulus  we calculated the stray-light contamination produced by  a uniform source extended all around the
field of view. To do this we linearly interpolated the calibration observations  in the 15\arcmin $< \theta  < $120\arcmin
interval  and integrated  fraz$_i$($\theta$) 2$\pi$ $\theta$ d$\theta$  (lower-left panel of Fig.~\ref{fig:slcom}).
In this way we get the contamination measure in terms of CXRB fraction.
The histogram  of the relative errors of  the 12 bins are plotted  in the lower right hand panel: typical value is 5\%. 

\noindent After summing different contributions at different off-axis angles and  accounting for the effective exposure map of 
our observation, we found that  in our ROI the stray-light contamination is expected to be
equivalent to the 0.690$\pm$0.020\% of the CXRB  or 
(4.9$\pm0.2$) $\times$10$^{-12}$ erg s$^{-1}$ cm$^{-2}$  deg$^{-2}$ in the 2.0-10.0 keV.
This is done in the assumption that the CXRB all around the CDFS is uniform and accurately reproducible.
The first assumption is justified by the fact that  the master ROSAT catalogs 
\footnote{\texttt{http://www.xray.mpe.mpg.de/$\sim$jer/rosat/cats/catcat.html}}
do not report bright sources  within a two degree radius. This means that we can safely assume that the contamination 
on our field of view is produced by a typical collection of CXRB sources that are symmetrically distributed.  
It also follows that  the second assumption is fully justified. In fact, although we do not have any direct  X-ray observation 
of the nearby regions (ROSAT observation are limited to energies $<$ 2keV),  the cosmic variance is  $\lesssim$ 3\%
because the interested sky area is as large as $\sim$3 deg$^2$ \citep{Moretti09}.
These uncertainties are accounted for in the systematics calculation, see Sect~\ref{sect:syst} .

\section{Fit procedure and results} \label{sect:procures}
\begin{figure}
\includegraphics[width=9cm] {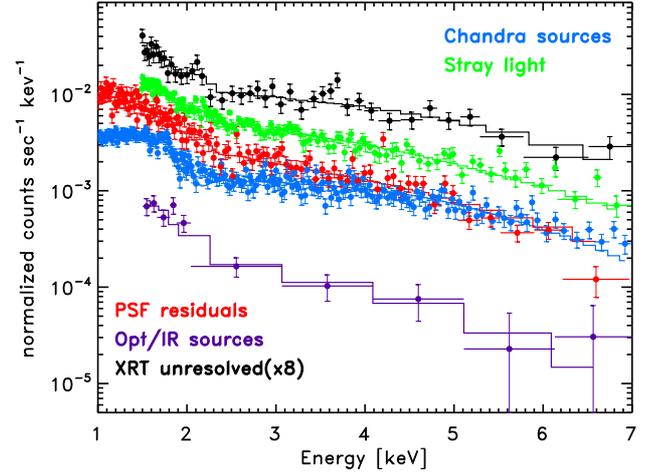}      
\caption{The five data sets we used  in this work. The XRT unresolved spectrum (black) has been fit by a  sum of 5 
different power law: (i) Chandra sources, not detected by XRT (blue); (ii) X-ray emission of optical/IR sources (violet) ; 
(iii) XRT sources to model PSF residuals (red); (iv) the CXRB spectrum to model stray-light contamination (green). 
The XRT unresolved spectrum has been offset up by a factor 8 for clarity. }
\label{fig:5spec}
\end{figure}
We used a total of five different datasets (Fig.~\ref{fig:5spec}), plus one for the instrument background:
(i) the Chandra stacked spectrum of the 326 sources undetected by Swift-XRT; (ii) the Chandra stacked spectrum of the 5683 optical/IR sources; 
(iii) the XRT stacked spectrum of the 109 detected sources, used with the ad-hoc modified 
response file to model the PSF residuals; (iv) the CXRB spectrum (stacking of 130 XRT observations) used to quantify the stray-light contamination; 
(v) the fifth dataset is the XRT unresolved spectrum that we assume is
given by the sum of the four other (opportunely renormalized), plus the truly unresolved X-ray emission.
To correctly propagate the statistical uncertainties of the single components, we fitted  the five datasets together.
We modeled (i)-(iv) datasets with single absorbed power laws. 
For the dataset (v) we used the sum of five different power laws (wabs*(pow+pow+pow+pow+pow) in XSPEC syntax):
the parameters of the first four power-laws were linked to the best values of the (i)-(iv) dataset fit,  while
the fifth photon index and normalization were left  free to vary. The absorbing column
was frozen to the Galactic value derived from the HI Galaxy map  \citep{Kalberla05} 7.02$\times$10$^{19}$cm$^{-2}$. 
\begin{figure}
\includegraphics[width=9cm] {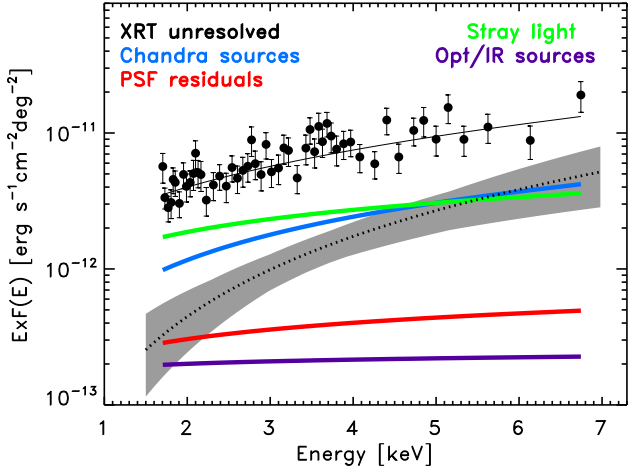}      
\caption{Results of the fit procedure. We modeled the XRT unresolved spectrum as the sum of five components, each fit by a power law.  
Colors are the same as in the previous figure: (i) Chandra sources not detected  by XRT (blue); x
(iii) Bright sources PSF residuals (red); (iv) stray-light contamination (green);
(iv) X-ray emission of optical/IR sources (violet) . The fifth component, the 
truly unresolved emission, which is our aim, is plotted with its uncertainties as the gray area.}
\label{fig:colspec}
\end{figure}
\noindent In the fit procedure we used C-statistics grouping the spectrum to have a minimum of one count for each bin. We checked that enhancing the minimum
event number for each bin (we tested 3 and 5) does not significantly affect the fit results. 
We also checked that grouping data with a minimum of 20 counts for each bin and using chi-square statistics does not significantly change the results.

\subsection{Results}
\label{sect:results}
\noindent Considering the five data-sets together, the adopted model provides a good description of the data:
chi squared = 2260.25 using 1954 PHA bins; reduced chi-squared = 1.16 for 1944 degrees of freedom. The best values 
are reported in Tab.~\ref{tab:contrib}, and the best-fit models are plotted in Fig.~\ref{fig:colspec}.
Instrument background and the stray-light contamination represent $\sim$ 52\% 
and $15$\% of the signal respectively. The remaining 33\% is essentially contributed by the emission of the Chandra sources
and  the real unresolved emission at the same level (15\% each), while the optical sources represent 2\% of the total. 

\noindent We find that the unresolved emission can be modeled by a hard power law 
with photon index $\Gamma$=0.1$\pm$0.7 and a normalization of 7.6 (+10.0,-5.2) $\times$10$^{-5}$ photons
s$^{-1}$ cm$^{-2}$ deg$^{-2}$ at 1 keV, corresponding to a flux density 
of 5 $\times$10$^{-12}$ erg s$^{-1}$ cm$^{-2}$ deg$^{-2}$  in the 2.0-10. keV band.
\begin{figure}
\includegraphics[width=8cm] {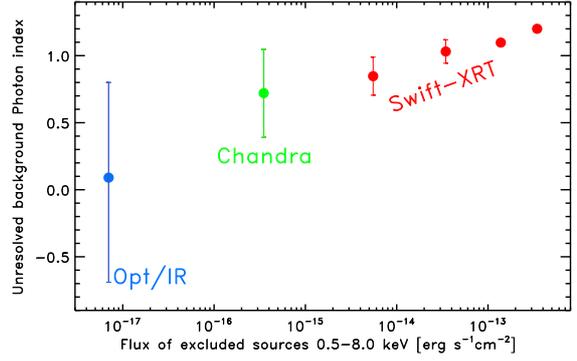}      
\caption{The photon index of the unresolved CXRB as function of the flux of the excluded sources.}
\label{fig:harden}
\end{figure}
We note that excluding fainter and fainter sources makes the unresolved spectrum harder. As shown in Fig.~\ref{fig:harden}, 
considering the whole signal from the CDF-S, including all the sources,  yields a  photon index is 1.20$\pm$0.05, which is close to
the total CXRB ($\Gamma$=1.4). Excluding all the XRT detected source makes the slope 0.84$\pm$ 0.1, while excluding all the
Chandra 4 Ms catalog sources brings the photon index down to 0.7 $\pm$ 0.3.  Finally removing the contribution of optical/IR
sources makes  the final  $\Gamma$=0.1$\pm$0.7.

\noindent In the remainder of this section we detail the results we obtain for the four other components of the XRT unresolved signal. 

\noindent (i) We find that the integrated emission of the 326 Chandra sources is well constrained 
both in slope (0.94$^{+0.03}_{-0.03}$) and flux (4.94$^{+0.12}_{-0.12}$ $\times$10$^{-12}$  erg s$^{-1}$ cm$^{-2}$ in the 2-10 keV band). 
As expected, the total integrated emission of these faint sources ( flux $<$ 10$^{-16}$ and 10$^{-15}$ erg s$^{-1}$ cm$^{-2}$  
in the 0.5-2.0 and 2-10 keV bands respectively) is harder than the spectrum of brighter sources. 
\begin{table}
\caption[]{Total flux of the 326 Chandra sources not detected in the XRT image: comparison of our best fit to the stacked spectrum
with the sum of the \cite{Xue11} catalog photometric values.}
\begin{tabular}{|l|cc|}
\hline
                  & flux$_{\rm 0.5-2.0 keV}$        & flux$_{\rm 2.0-8.0 keV}$           \\ 
                  & [erg s$^{-1}$ cm$^{-2}$]  & [erg s$^{-1}$ cm$^{-2}$]      \\   
\hline
Present work (spec.)    & 2.69$^{+0.06}_{-0.06} \times$10$^{-14}$    &  1.12$^{+0.03}_{-0.03} \times$10$^{-13}$   \\
&&\\
\cite{Xue11}  (photo.)   & 2.70$^{+0.05}_{-0.05} \times$10$^{-14}$    &  1.05$^{+0.07}_{-0.07} \times$10$^{-13}$   \\
\hline
\end{tabular}
\label{tab:chacheck} 
\end{table}
As a check we compared  these fluxes with the photometry of  \cite{Xue11} catalog, which is reported in the 
0.5-2.0 keV and 2.0-8.0 keV bands. 
We find very good agreement between the two measures both  in the soft and hard bands (Tab.~\ref{tab:chacheck}).

\noindent (ii) Considering the sources detected in XRT to calculate the PSF residuals, we get a photon index of 1.61$^{+0.05}_{-0.05}$. 

\noindent (iii) The best-fit slope of the  130 XRT stacked observations (to quantify the stray-light contamination) is1.46$^{+0.05}_{-0.04}$.

\begin{figure}
\includegraphics[width=8cm] {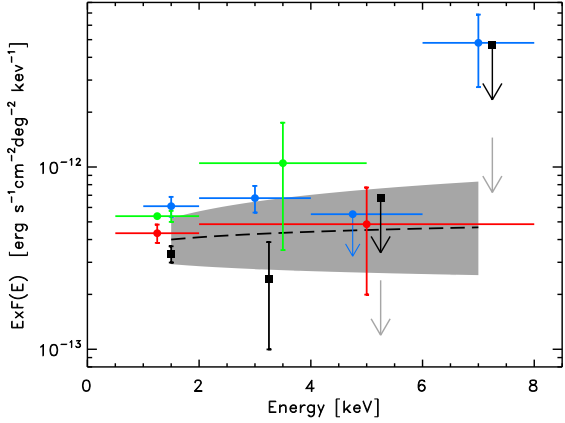}      
\caption{Comparison among different measures of the contribution of optical/IR (X-ray undetected) to the X-ray background. 
The gray area represents the best fit to our data. Green, blue,  and red points represent the photometry of 
\cite{Hickox07}, \cite{Luo11}, and \cite{Cowie12}. Black squares and arrows represent our photometric values and
2$\sigma$ upper limits  including  systematic error of 2\% on the background measure. 
Gray arrows show the upper limits accounting only for statistical errors.}
\label{fig:optflux}
\end{figure}

\noindent (iv) The X-ray emission of the optical/IR sources can be modeled well by a relatively soft power law (photon index 1.9) and a flux of  
7.2$^{+2.6}_{-1.9}$ $\times$10$^{-13}$  erg s$^{-1}$ cm$^{-2}$  deg$^{-2}$ in the 2-10 keV band.
These values are in very good agreement  with \cite{Cowie12}, who use the same optical/IR catalog and report a 
flux of  $\sim$ 7.0   $\times$10$^{-13}$  erg s$^{-1}$ cm$^{-2}$  deg$^{-2}$ in the 2.0-10.0 keV band, 
(calculated from their Tab.~1) and a photon index $\Gamma$=1.7-2.0 based on X-ray colors (Fig.~\ref{fig:optflux}).
Our measure is slightly lower but consistent with \cite{Hickox07} and \cite{Luo11} (up to 6 keV) who use different 
catalogs. 
 In the hard part of the spectrum, we cannot  confirm the results of   \cite{Luo11}  who find 
 a strong signal  equal to 4.8$\pm$2.0 erg s$^{-1}$ cm$^{-2}$  deg$^{-2}$ (28\%$\pm$10.0 of the total CXRB)
 in the 6-8 keV band. While their measure is in good agreement with our spectroscopy in the 0.5-6.0 band, consistent 
 with a photon index $\Gamma$=2, their photometry  is  an order of magnitude higher than our spectrum  in the 6-8 keV interval.
They interpret this strong feature as the signature of a substantial population of highly absorbed AGN \citep{Xue12}. 
The discrepancy with our data cannot be explained by the different optical/IR adopted catalogs:  
indeed with their catalog  \citep{Giavalisco04}  we find results that are fully consistent 
with our previous finding both in spectral slope and in flux. 
In order to go into tho discrepancy thoroughly, we also performed the photometry 
in the [0.5-2.0],[2.0-4.0],[4.0-6.0], and [6.0-8.0] keV bands.\footnote{We calculated the signal registered within the 1\arcsec~radius circular regions
centered on the positions of the 8637 sources listed by the \cite{Giavalisco04} catalog within the 4.2\arcmin radius 
circular region shown in Fig.~\ref{fig:optfov}, in the [0.5-2.0],[2.0-4.0],[4.0-6.0],[6.0-8.0] keV bands. 
To estimate the background we use the mean flux recorded in the same region after  excluding of both the X-ray and 
optical/IR sources. In the 6-8 keV we count 25,385 total events with 25,331 expected background events over a total
of 103973 pixels.  To convert these counts in flux, we use a mean exposure time of  3296249.2 seconds  and  a conversion factor
of 1 count s$^{-1}$ = 7.1$\times$10$^{-11}$erg  s$^{-1}$ cm$^{-2}$. 
Using a larger aperture for the source extraction regions we find consistent results. }
 While we find agreement with all previous works up to 6 keV,  we do not find any significant flux beyond 4 keV (Fig.\ref{fig:optflux}).
Indeed, allowing only for statistical uncertainties our photometry in the 6-8 keV band yields 
an upper limit that is inconsistent with the flux measured by \cite{Luo11}.
However, this discrepancy can be explained by a slight difference in background estimate.
Indeed allowing for only a 2\% systematic error in our background value, we find that 
our 2$\sigma$ upper limit becomes consistent with the photometric point by \cite{Luo11}.
This is because, once all the bright sources have been removed, in the 6-8 keV, 
where a strong instrument feature is also present, the ratio between the instrument 
background and the cosmic signal is $\lesssim$ 1\%  (see for example http://cxc.harvard.edu/contrib/maxim/stowed/).

\begin{table*}
\begin{tabular}{|l|c|cc|c|}
\hline
                                &    slope   &   flux [erg s$^{-1}$ cm$^{-2}$ deg$^{-2}$ ]  & flux [erg s$^{-1}$ cm$^{-2}$ deg$^{-2}$ ]& fract.[\%] \\  
                                &            &   (1.0-2.0) keV                              & (2.0-10.0) keV                           & (1.5-7.0)keV  \\  
\hline
 Chandra sou. & 0.94     (-0.03,+0.03)&5.70e-13     (-1.39e-14,+6.63e-15)&4.94e-12     (-1.20e-13,+1.23e-13 )& 14.7\\
    PSF res.  & 1.61     (-0.05,+0.05)&1.84e-13     (-7.53e-15,+5.74e-15)&6.88e-13     (-2.81e-14,+2.90e-14 )&  2.0\\
 S.L. contam. & 1.46     (-0.04,+0.05)&1.08e-12     (-3.95e-14,+3.00e-14)&4.79e-12     (-1.75e-13,+1.98e-13 )& 14.3\\
  OPT/IR sou. & 1.90     (-0.37,+0.45)&2.76e-13     (-7.18e-14,+8.94e-14)&7.25e-13     (-1.88e-13,+2.58e-13 )&  2.2\\
\hline
\hline
   Unresolved & 0.09     (-0.78,+0.70)&1.77e-13     (-1.16e-13,+1.51e-13)&5.00e-12     (-2.62e-12,+3.22e-12 )& 14.9\\

\hline
\end{tabular}
\caption[]{Best fit values of the different  components contributing the unresolved signal in the XRT observation. 
Errors calculated at 68\% confidence for two  parameters ($\Delta\chi^2$=2.3). In the last line we report the power-law best-fit 
values to the extragalactic unresolved background.  
The reported errors are the quadratic sum of statistic and systematic error. The last column reports the relative contribution of the 
single components, taking into account that the instrument background is 53\% of the total. }
\label{tab:contrib} 
\end{table*}
%

\subsection{Systematic  errors} \label{sect:syst}
\noindent Because we work in a very low signal/background regime (15\%) in the uncertainties calculation we considered the impact 
of the systematics of the single components of the model. To do this we repeat the fit procedure 1,000 times, each time varying 
the normalization of the different components according to Gaussian distributions with the variances reported above: 5\% for the instrument background, 
7\% for the Chandra source spectrum, 5\% for the stray light, and 10 \% for PSF residuals. 
We found that the impact of these systematics on the final slope uncertainty is modest, 0.08 (1$\sigma$), which is
only 10\% of the statistical error. The impact on the flux uncertainty is greater at 1.77 erg s$^{-1}$ cm$^{-2}$ deg$^{-2}$,
which is 75\% of the statistical error. 

\section{Discussion}\label{sect:disc} 
\subsection{Comparison with previous works}
\begin{figure}
\includegraphics[width=9cm] {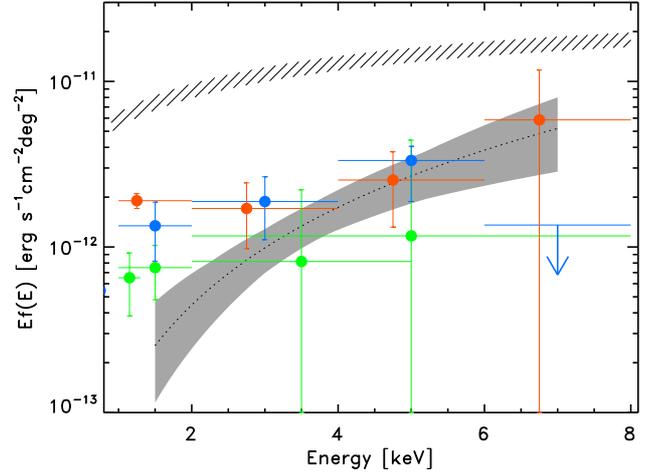}      
\caption{The spectrum of the unresolved CXRB: comparison with previous measures. The gray area is the result  of the present work. 
Green points are from \cite{Hickox07}; blue points are from \cite{Luo11}, with  the blue arrow indicating the 1$\sigma$ upper limit; 
red points from \cite{Worsley06}. 
The dashed area represents the total CXRB emission as measured by Swift \citep{Moretti09}.} 
\label{fig:fraz}
\end{figure}

%
The unresolved emission has been directly measured at faintest level by
\cite{Hickox07} who found that in both the 
CDF-S and CDF-N it was consistent with zero, although with large uncertainties: in the CDF-S they measured 1.4$\pm$3.9 $\times$10$^{-12}$ erg s$^{-1}$ cm$^{-2}$ deg$^{-2}$ in the band 2-8 keV. 
This is nearly a factor three smaller than our measurement, but still consistent within 1$\sigma$ (Fig.~\ref{fig:fraz}).
\noindent 
One could find surprising it that the level of unresolved emission we measure in the 4Ms is three times higher than 
the one found by a previous work  based on 2Ms. 
The reason for this is that the sources in the 4 Ms observation that are not detected in 2 Ms supply only a small fraction 
of the total CXRB flux.
Indeed, while there are many sources of the 4 Ms catalog \citep{Xue11} that are not included in the 2Ms catalog by \cite{Luo08}
(169 out of 430 in our ROI, that is 40\%), their total flux is only $\sim$10$^{-14}$ erg s$^{-1}$ cm$^{-2}$,
which represents only 1.5\% of the total CXRB or $\sim$  7\% of the unresolved.  
However we stress that  our measure represents an improvement in the sense that the 1$\sigma$ upper limit 
(measured value + 1$\sigma$ error) is lower than the previous ones.

\noindent   In the soft band we find that the spectral fit between 1.5 and 2.0 keV is only marginally consistent 
with the  \cite{Hickox07} photometry in the band 0.5 -2.0.
However, we note  that a fair comparison between the two measures is not possible because the energy bands 
do not completely coincide. Moreover, in the 0.5-1.5 keV band the spectrum of the unresolved  background 
is expected to have a thermal component owing to  undetected galaxy groups and WHIM emission. This means 
that in the soft band the extrapolation of our best fit probably underestimates the CDFS unresolved  background.
As already said we limit our measure to energies higher than 1.5 keV because our data do not allow us to 
accurately disentangle these thermal extragalactic  components from the local thermal emission of 
the Galaxy and Local Bubble.  

\noindent In Fig.~\ref{fig:fraz} we also plot the results of \cite{Worsley06} and \cite{Luo11}, which are in good agreement 
with our results in the  2-6 keV band. 
For the reasons already discussed our spectrum is not directly comparable to  their photometry in the 0.5 -2.0 band. 

\cite{Luo11} find that in the 6-8 keV the unresolved emission is consistent with zero with an upper limit at 10\%
of the total CXRB at 7 keV.  This is essentially  due to the measure of the contribution of the optical and infrared 
sources, which  they find at the level of 28\% of the total CXRB in this particular 
energy interval (see Sect~\ref{sect:results}). 
At odds with this measure, we do not find any significant feature in the X-ray  spectrum of optically/IR
detected sources, resulting in a estimate of the unresolved flux at 7 keV which is four times higher than  \cite{Luo11}  1 $\sigma$
upper limit. 
    
\noindent 
In terms of the resolved fraction of the total CXRB, using the Swift measurement of the total CXRB \citep{Moretti09},  
our measurement of the unresolved components corresponds to $\sim$80\% in the 2-10 keV 
band, with 95$\pm$3\% at 2 keV and 70$\pm$16\% at 7 keV (lower panel of Fig.~\ref{fig:fraz}). 
We note that this number should be kept with caution and does not really add any significant piece of news
once the unresolved emission is directly probed and constrained.  
This is not only because the spectrum of the total CXRB is highly uncertain \citep{Revnivtsev05, Frontera07, Moretti09},
but also because deep fields are limited to a small portion of the sky so they miss
the rare and bright population that is responsible for $\sim$20\% of the total emission.

\noindent
To summarize, we can state that, compared  to previous works and thanks to the low level of the Swift-XRT instrument background in 
the hard band ($>$2 keV) we significantly reduced the uncertainties finding consistent results in terms of absolute flux measurement.
After reducing the uncertainties our result left room for a small, but significant component still to be resolved.
%
%
\subsection{Comparison with AGN population models}
\begin{figure*}
\includegraphics[width=9cm] {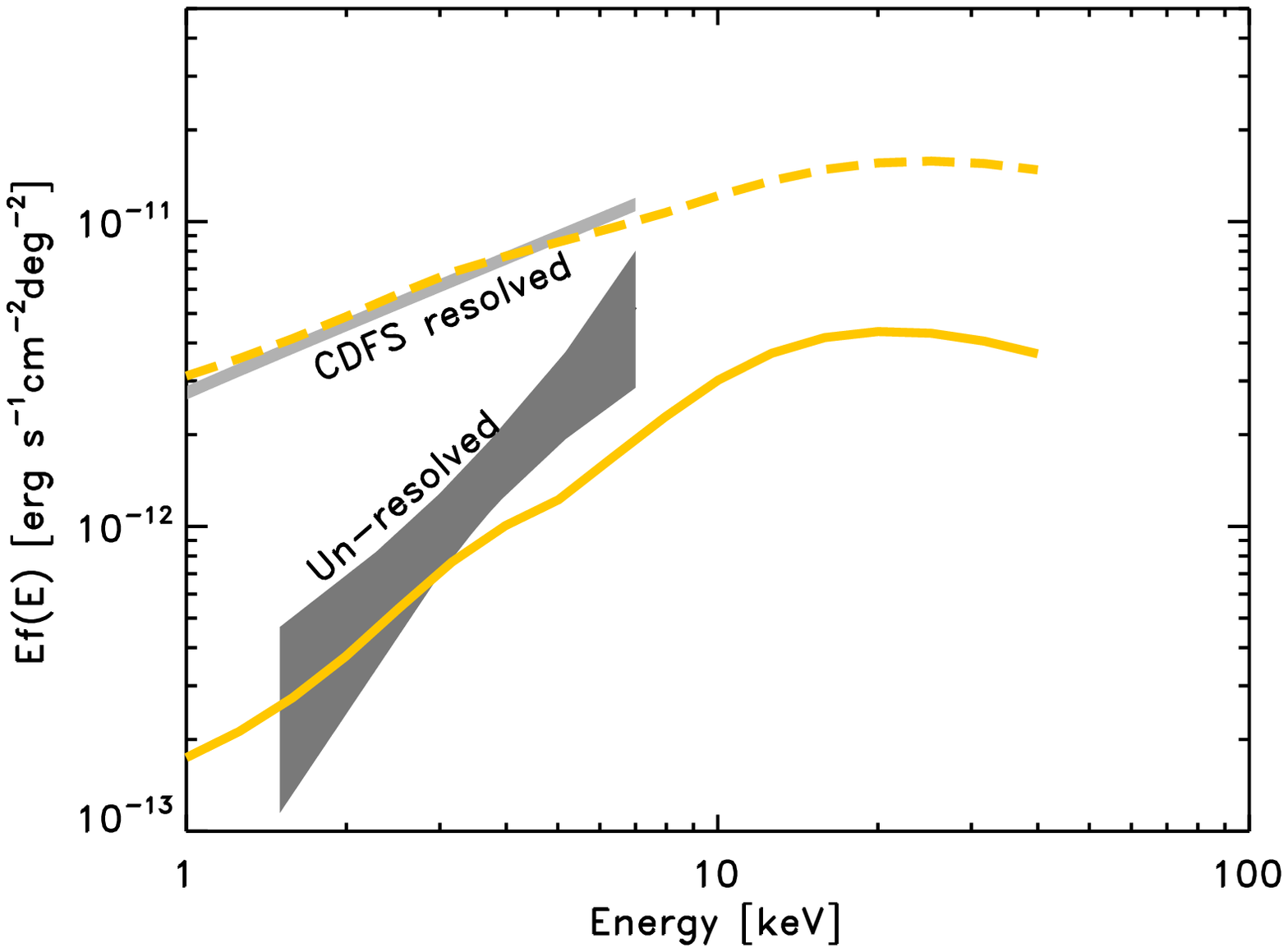}  
\includegraphics[width=9cm] {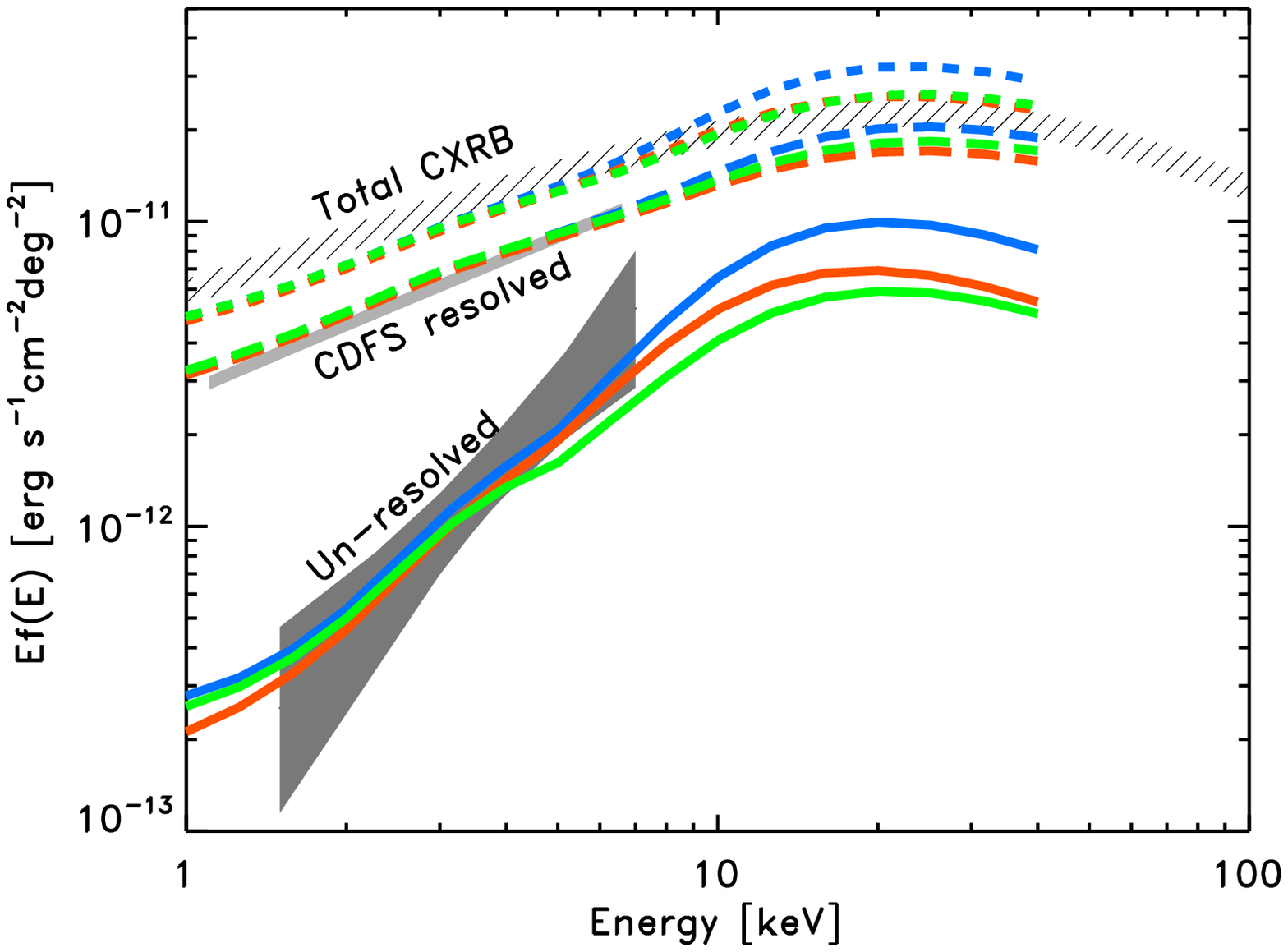}     
\caption{In both panels the dark and light gray areas represent the spectrum of the unresolved and resolved CXRB, respectively. 
{\bf Left Panel:} The yellow dashed and continuous lines show the expectations of the G07 model for the two components. 
{\bf Right Panel:}
The shaded area represents the total CXRB emission. Blue lines show the G07 model output with the CT (any z, any L, nH $>$ 24)
contribution enhanced by a factor 2. Green lines show the G07 model with the only heavily CT contribution enhanced by a factor 5. 
Red lines show the case where the contribution of high z CT has been enhanced
assuming a linear evolution with redshift. Continuous, long-dashed, and short-dashed lines are for CDFS unresolved, CDFS resolved, 
and total CXRB, respectively.}
\label{fig:pompa}
\end{figure*}

\noindent To check whether the  unresolved spectrum can be accounted for by AGN population synthesis
models, we compared our results with \cite{Gilli07} (G07) models \footnote{\texttt{http://www.bo.astro.it/$\sim$gilli/xrb.html}},
which provides the integrated spectrum for different AGN subpopulations at given luminosity (L), redshift (z) and absorbing 
column density (N$_{\rm H}$). 

\noindent To compare data and model, first of all, we need to assess which AGN subpopulation can be detected in the CDFS observation. 
Therefore, for a grid of six n$_{\rm H}$ (10$^{20}$-10$^{26}$ cm$^{-2}$),   thirty-six L (10$^{41}$-10$^{48}$ erg s$^{-1}$ cm$^{-2}$), and twelve z (0-6), 
we calculated the expected median flux, using the same spectral models as used in G07.
For each bin flux we calculated the detection probability by means of the response function of the \cite{Xue11} catalog.
The \cite{Xue11} catalog is the result of the cross-correlation of three different catalogs built in three different
energy bands, while the response function is only provided separately. We calculated the detection probability 
as the combination of the probability in the soft and hard bands. 
To ease the calculation we neglected the possibility that a source can only be detected in the total band, while it is undetected 
either in the soft or in the hard. In our ROI this is only true for 12 out of 430 sources (2.7\%).

\noindent For example, considering AGN with photon index $\Gamma$=1.9 in the bin 
10$^{22}<$n$_{\rm H}<$10$^{23}$, 1.0$<$z$<1.5$ and 10$^{41.9}<$L$<$10$^{42.1}$ erg s$^{-1}$ cm$^{-2}$,
we expect fluxes of 6.0 $\times$ 10$^{-17}$  and 1.8 $\times$10$^{-16}$ erg s$^{-1}$ cm$^{-2}$  in the 0.5-2.0,  and 2.0-8.0 bands, respectively.
The catalog response function (Table 7 and Fig. 23 of  \cite{Xue11})  gives the source detection probability at these fluxes  
in the two bands we are interested in,  p$_{soft}$=0.4 and p$_{hard}$=0.2.
We calculate that, for this AGN,  the probability to be included in the CDF-S catalog is p$_{det}$=1.0- (1-p$_{soft}$)$\times$(1-p$_{hard}$), 
that is 0.52.

\noindent As in G07 a distribution of photon index was considered,  we repeated the same procedure using nine different slopes  (between 1.5 and 2.3)
and weighting the final output by the same distribution used by G07 that is a  Gaussian centered on 1.9 with standard deviation 0.2 .

\noindent The G07 model is fully consistent (7\% scatter) with the summed spectrum of all the  
in  \cite{Xue11} catalog  (Fig.~\ref{fig:pompa}). We also find that the same model  accurately reproduces the soft part 
($\lesssim$ 3 keV) of the spectrum of the unresolved emission, whereas it falls short in replicating the 3-7 keV emission, 
hinting that there are some missing hard sources. 

\subsubsection{AGN and galaxies}
\noindent To compare our data with the G07 model we assumed that the emission of the 5683 optical/IR sources from
the  \cite{Grazian06} 
catalog, subtracted from the unresolved spectrum, is dominated by non-nuclear sources, with a negligible contribution 
from an AGN component. There is some evidence of this hypothesis. 
First, although we cannot measure the X-ray fluxes of the single sources (which are by definition fainter than the detection limit), 
we can estimate the mean luminosity of the sample, using the cataloged photometric redshifts. We find that a mean luminosity 
of 2$\times$10$^{39}$ erg s$^{-1}$ cm$^{-2}$ in the 0.5-2.0 keV band accounts for the measured total flux of the sample
(7.25$_{-1.88}^{+2.58}\times$10$^{-13}$ erg s$^{-1}$ cm$^{-2}$ deg$^{-2}$,see Table~\ref{tab:contrib}).  
Such faint mean luminosity, together with the softness of the spectrum (photon index $\sim$ 1.9), is a clear indication that the galactic 
emission is prevalent with respect to the nuclear component.
While the mean luminosity by itself can be an indication, it does not rule out the presence of some AGN.
However, integrating (the extrapolation of) the X-ray galaxy number-counts distribution measured by \cite{Lehmer12}
in the 2.-10 keV band down to the flux (3$\times$10$^{-19}$ erg s$^{-1}$ cm$^{-2}$) for which the source density corresponds to 
the observed ($\sim$ 400,000 galaxies per deg$^2$) and accounting for the catalog completeness function, we find a value of 
6.5 10$^{-13}$ erg s$^{-1}$ cm$^{-2}$ deg$^{-2}$, which is very close to and consistent  with our measure.  
Moreover, a completely independent estimate of the galaxy X-ray flux density comes from  \cite{Dijkstra12} who
estimate the contribution of the galaxies to the CXRB. 
Lacking an accurate measure of the mean galaxy spectral slope in the hard band, they calculated this contribution 
by assuming different mean photon indices  and using the star formation history of \cite{Hopkins06} together with the star formation rate 
to X-ray luminosity conversion of \cite{Mineo12b}.  
This estimate is particularly suitable for our purposes as we have a direct measure of the galaxy mean spectral slope.
In the case of $\Gamma$=1.9, they estimated that the galaxy density flux is 1.2$\pm$0.3$\times$10$^{-12}$ erg s$^{-1}$ cm$^{-2}$deg$^{-2}$ 
in the 2-10 keV band, which is slightly higher than our measure and perfectly consistent with our assumption.
These arguments suggest that the emission of the X-ray undetected  optical/IR sources of 
\cite{Grazian06}  catalog is mostly of galactic origin with a negligible nuclear contribution. 

%
\subsection{Some speculations}
\noindent We examine the possibility that the discrepancy  between the unresolved spectrum and the G07 model could
be explained by the cumulative emission of a number of Compton-thick (CT) which are not included in the model.    
From an observational point of view the CT luminosity function is 
highly uncertain both in the local Universe  \citep{DellaCeca08, Treister09, Ballantyne11, Severgnini12, Ajello12} and at higher redshift 
\citep{Tozzi06, Daddi07, Alexander08, Fiore09, Alexander11}. 
In G07 the  CT luminosity functions at different redshift are indirectly calculated from the difference between the 
total CXRB and the integrated emission of  Compton-thin  AGN (21$<$N$_{\rm H}<$ 24), under the assumption that the number of mildly 
CT objects (log N$_{\rm H}$ = 24.5) is equal to that of heavily CT objects (log N$_{\rm H}$ = 25.5) and that they have 
the same cosmological evolution of Compton-thin AGN. 

\noindent  While the total integrated emission of the CT population by definition agrees with the total CXRB,  
as we said in the previous section, we found some discrepancies with the constraints placed by the unresolved emission spectrum.
Multiplying the whole CT contribution (at any redshift and luminosity) by a factor 2.0 would reconcile model 
and the observation in the 1.5-7 keV band, but, at the same time, would break through the spectrum 
of the total CXRB at harder energies $\sim$ 20-30  keV, where the CT contribution is less affected by absorption 
(blue line in the right panel of Fig.~\ref{fig:pompa}).

\noindent Enhancing the contribution of the only heavily CT population by a factor five significantly reduces the distance between 
data and model; but the number of these sources, which is required to make the model consistent with the data under 7 keV, would 
result in over predicting the total CXRB at higher energy (green line in right panel of Fig.~\ref{fig:pompa}). 

\noindent On the other hand, high-redshift (z $>$2) CT are expected to peak at lower energies and to supply
less flux at energies $>$ 20 keV.  Indeed, while in G07 model a constant fraction of obscured AGN with redshift is assumed, 
we find that, hypothesizing a linear positive evolution with redshift (1+z) in the total number of CT objects, corresponding to 
a  (1+z)$^a$ with a$\sim$ 0.5 in terms of CT fraction, would solve the discrepancy between model and observation. 
With this assumption  the overall AGN contribution is still consistent with the CXRB constrain at higher energies 
(red line in the right panel of Fig.~\ref{fig:pompa}).
\begin{figure}
\includegraphics[width=9cm] {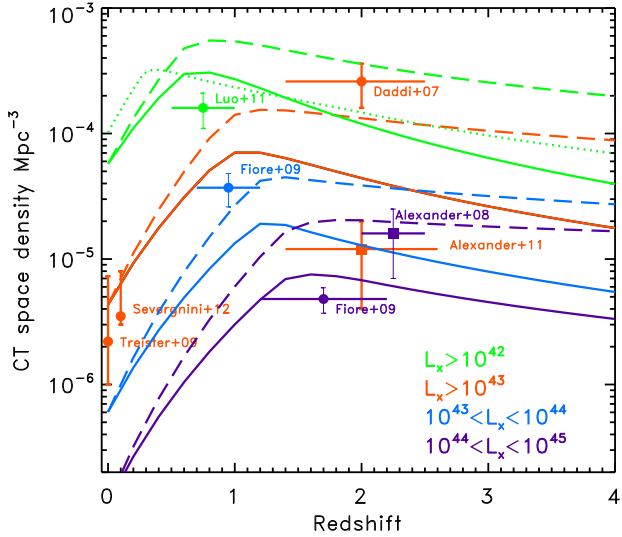}      
\caption{The comoving space density of the CT AGN. We report different results for different redshift and luminosity intervals.
Different colors indicate different luminosity bins. Continuous lines show the output of the G07 model; dashed lines represent the same outputs
modified by applying a linear positive evolution in redshift. The green dotted line displays the output of the composite model by \cite{Draper10} }
\label{fig:zevol}
\end{figure}

\noindent In Fig.~\ref{fig:zevol} we report different measurements of the CT space density performed with different techniques 
(most in the infrared band) at different redshift, for different X-ray luminosity intervals, together with the output of the G07 model and 
imposing a linear evolution as suggested by our observation. This shows that such an evolution of the absolute number of
CT AGN is still consistent with the large scatter of the observation results.

\noindent Our solution would require an increase in the CT fraction with redshift. 
The change in the obscured AGN fraction with redshift is a debated issue.
A positive evolution has been observed in some studies \citep{Lafranca05,Treister06, Ballantyne06, Hasinger08}, whereas
others do not find any significant variation \citep{Ueda03,Akylas06}.
While these studies adopted a  value of $\sim$10$^{22}$cm$^{-2}$ as pivotal column density 
to separate obscured and unobscured sources, our data seem to indicate a stronger evolution for only the CT population
($>$10$^{24}$cm$^{-2}$ ).   

\noindent
A similar result was found by \cite{Brightman12} that shows
a significant increase in the only CT fraction, from 20\% in the local universe up to 40\%  at z=1-4. 
An higher number of high-redshift CT have recently been claimed by \cite{Gilli11} who find a CT AGN 
at z=4.76 in the CDF-S area, the detection probability being less than one.   
A strong positive evolution of the only CT population, has been also 
discussed in \cite{Treister09} as a possible explanation to reconcile the difference between IR observations 
and the expectations from AGN population models normalized to reproduce the local CT density.

\noindent From the viewpoint of the population modeling,  evolution of the only CT population 
has been found by \cite{Draper10} as a natural consequence of describing the CT AGN  by means of a
  physically motivated  Eddington ratio distribution: CT AGN are a composite population accreting either at $>$90\% or $<$1\%.
Indeed in this model the predicted evolution of the number density of CT AGN seems to follow the one we propose,
although the normalization  is a factor 2 lower (Fig.~\ref{fig:zevol}).

\noindent 
In the framework of AGN-galaxy coevolution, a positive redshift evolution of heavily
obscured AGN is in line with the idea that in the high-redshift Universe
SMBH are hosted by gas-rich galaxies. In particular, one of the scenario
proposed for the formation and evolution of SMBH and galaxy assumes that major
merger can trigger star formation and black hole growth at the same time.
Depending on the kind of interaction,  different
obscuration geometries could be possible around the black hole, but in any case 
this would increase the probability of observing a SMBH  through a very large amount of gas and dust.

\section{Conclusions}
\noindent We exploited the low and predictable instrument background of the Swift XRT telescope to 
perform the spectroscopy of the unresolved X-ray emission in the CDF-S.
We found a  faint, but significant unresolved component that  
can be modeled by a very hard power-law with photon index $\Gamma$=0.1$\pm$0.7 and a flux density 
of 5 $\times$10$^{-12}$ erg s$^{-1}$ cm$^{-2}$ deg$^{-2}$  in the 2.0-10. keV band, corresponding to 20\% of the total CXRB. 
With respect to previous works we significantly improved the accuracy over the 1.5-7.0 keV band. 
Our measure is in very good agreement with what is expected by the G07 AGN population model
in the 1.5-3.0 keV. In the hard band (3-7 keV) the same model falls short when replicating the observed spectrum,
pointing toward some missing  very hard sources.
This discrepancy can be solved hypothesizing a positive evolution with redshift
of the contribution Compton-thick AGN population. 

\begin{acknowledgements}
This work has been  supported  by ASI grants  I/011/07/0 and  ASI/INAF I/009/10/0.
and has made use of 

\noindent -- The XRT Data Analysis Software (XRTDAS) developed under the responsibility of the ASI
Science Data Center (ASDC), Italy;

\noindent -- the NASA's Astrophysics Data System;

\noindent -- the NASA/IPAC Extragalactic Database (NED) which is operated by the Jet Propulsion 
Laboratory, California Institute of Technology, under contract with the National Aeronautics and Space Administration.

We thank the anonymous referee for helpful comments that improved this work. AM is grateful to S. Andreon for useful
discussions on measure errors handling.

\end{acknowledgements}
\bibliographystyle{aa}   
\bibliography{tot09}   

\begin{thebibliography}{66}
\expandafter\ifx\csname natexlab\endcsname\relax\def\natexlab#1{#1}\fi

\bibitem[{{Ajello} {et~al.}(2012){Ajello}, {Alexander}, {Greiner}, {Madejski},
  {Gehrels}, \& {Burlon}}]{Ajello12}
{Ajello}, M., {Alexander}, D.~M., {Greiner}, J., {et~al.} 2012, \apj, 749, 21

\bibitem[{{Akylas} {et~al.}(2006){Akylas}, {Georgantopoulos}, {Georgakakis},
  {Kitsionas}, \& {Hatziminaoglou}}]{Akylas06}
{Akylas}, A., {Georgantopoulos}, I., {Georgakakis}, A., {Kitsionas}, S., \&
  {Hatziminaoglou}, E. 2006, \aap, 459, 693

\bibitem[{{Alexander} {et~al.}(2011){Alexander}, {Bauer}, {Brandt}, {Daddi},
  {Hickox}, {Lehmer}, {Luo}, {Xue}, {Young}, {Comastri}, {Del Moro}, {Fabian},
  {Gilli}, {Goulding}, {Mainieri}, {Mullaney}, {Paolillo}, {Rafferty},
  {Schneider}, {Shemmer}, \& {Vignali}}]{Alexander11}
{Alexander}, D.~M., {Bauer}, F.~E., {Brandt}, W.~N., {et~al.} 2011, \apj, 738,
  44

\bibitem[{{Alexander} {et~al.}(2008){Alexander}, {Chary}, {Pope}, {Bauer},
  {Brandt}, {Daddi}, {Dickinson}, {Elbaz}, \& {Reddy}}]{Alexander08}
{Alexander}, D.~M., {Chary}, R.-R., {Pope}, A., {et~al.} 2008, \apj, 687, 835

\bibitem[{{Ballantyne} {et~al.}(2011){Ballantyne}, {Draper}, {Madsen}, {Rigby},
  \& {Treister}}]{Ballantyne11}
{Ballantyne}, D.~R., {Draper}, A.~R., {Madsen}, K.~K., {Rigby}, J.~R., \&
  {Treister}, E. 2011, \apj, 736, 56

\bibitem[{{Ballantyne} {et~al.}(2006){Ballantyne}, {Shi}, {Rieke}, {Donley},
  {Papovich}, \& {Rigby}}]{Ballantyne06}
{Ballantyne}, D.~R., {Shi}, Y., {Rieke}, G.~H., {et~al.} 2006, \apj, 653, 1070

\bibitem[{{Bautz} {et~al.}(2009){Bautz}, {Miller}, {Sanders}, {Arnaud},
  {Mushotzky}, {Porter}, {Hayashida}, {Henry}, {Hughes}, {Kawaharada},
  {Makashima}, {Sato}, \& {Tamura}}]{Bautz09}
{Bautz}, M.~W., {Miller}, E.~D., {Sanders}, J.~S., {et~al.} 2009, \pasj, 61,
  1117

\bibitem[{{Brandt} \& {Hasinger}(2005)}]{Brandt05}
{Brandt}, W.~N. \& {Hasinger}, G. 2005, \araa, 43, 827

\bibitem[{{Brightman} \& {Ueda}(2012)}]{Brightman12}
{Brightman}, M. \& {Ueda}, Y. 2012, \mnras, 2850

\bibitem[{{Burrows} {et~al.}(2005){Burrows}, {Hill}, {Nousek}, {Kennea},
  {Wells}, {Osborne}, {Abbey}, {Beardmore}, {Mukerjee}, {Short}, {Chincarini},
  {Campana}, {Citterio}, {Moretti}, {Pagani}, {Tagliaferri}, {Giommi},
  {Capalbi}, {Tamburelli}, {Angelini}, {Cusumano}, {Br{\"a}uninger}, {Burkert},
  \& {Hartner}}]{Burrows05}
{Burrows}, D.~N., {Hill}, J.~E., {Nousek}, J.~A., {et~al.} 2005, Space Science
  Reviews, 120, 165

\bibitem[{{Cappelluti} {et~al.}(2012){Cappelluti}, {Ranalli}, {Roncarelli},
  {Arevalo}, {Comastri}, {Gilli}, {Rovilos}, {Vignali}, {Allevato},
  {Finoguenov}, {Miyaji}, {Nicastro}, {Georgantopoulos}, \&
  {Kashlinsky}}]{Cappelluti12}
{Cappelluti}, N., {Ranalli}, P., {Roncarelli}, M., {et~al.} 2012, ArXiv
  e-prints

\bibitem[{{Citterio} {et~al.}(1994){Citterio}, {Conconi}, {Ghigo}, {Loi},
  {Mazzoleni}, {Poretti}, {Conti}, {Mineo}, {Sacco}, {Braeuninger}, \&
  {Burkert}}]{Citterio94}
{Citterio}, O., {Conconi}, P., {Ghigo}, M., {et~al.} 1994, in SPIE Conference
  Series, ed. R.~B. {Hoover} \& A.~B. {Walker}, Vol. 2279, 480--492

\bibitem[{{Cowie} {et~al.}(2012){Cowie}, {Barger}, \& {Hasinger}}]{Cowie12}
{Cowie}, L.~L., {Barger}, A.~J., \& {Hasinger}, G. 2012, \apj, 748, 50

\bibitem[{{Daddi} {et~al.}(2007){Daddi}, {Alexander}, {Dickinson}, {Gilli},
  {Renzini}, {Elbaz}, {Cimatti}, {Chary}, {Frayer}, {Bauer}, {Brandt},
  {Giavalisco}, {Grogin}, {Huynh}, {Kurk}, {Mignoli}, {Morrison}, {Pope}, \&
  {Ravindranath}}]{Daddi07}
{Daddi}, E., {Alexander}, D.~M., {Dickinson}, M., {et~al.} 2007, \apj, 670, 173

\bibitem[{{De Luca} \& {Molendi}(2004)}]{Deluca04}
{De Luca}, A. \& {Molendi}, S. 2004, \aap, 419, 837

\bibitem[{{Della Ceca} {et~al.}(2008){Della Ceca}, {Caccianiga}, {Severgnini},
  {Maccacaro}, {Brunner}, {Carrera}, {Cocchia}, {Mateos}, {Page}, \&
  {Tedds}}]{DellaCeca08}
{Della Ceca}, R., {Caccianiga}, A., {Severgnini}, P., {et~al.} 2008, \aap, 487,
  119

\bibitem[{{Dijkstra} {et~al.}(2012){Dijkstra}, {Gilfanov}, {Loeb}, \&
  {Sunyaev}}]{Dijkstra12}
{Dijkstra}, M., {Gilfanov}, M., {Loeb}, A., \& {Sunyaev}, R. 2012, \mnras, 421,
  213

\bibitem[{{Dijkstra} {et~al.}(2004){Dijkstra}, {Haiman}, \&
  {Loeb}}]{Dijkstra04}
{Dijkstra}, M., {Haiman}, Z., \& {Loeb}, A. 2004, \apj, 613, 646

\bibitem[{{Draper} \& {Ballantyne}(2010)}]{Draper10}
{Draper}, A.~R. \& {Ballantyne}, D.~R. 2010, \apjl, 715, L99

\bibitem[{{Ettori} \& {Molendi}(2010)}]{Ettori10}
{Ettori}, S. \& {Molendi}, S. 2010, ArXiv e-prints

\bibitem[{{Fan} {et~al.}(2006){Fan}, {Strauss}, {Becker}, {White}, {Gunn},
  {Knapp}, {Richards}, {Schneider}, {Brinkmann}, \& {Fukugita}}]{Fan06}
{Fan}, X., {Strauss}, M.~A., {Becker}, R.~H., {et~al.} 2006, \aj, 132, 117

\bibitem[{{Fiore} {et~al.}(2009){Fiore}, {Puccetti}, {Brusa}, {Salvato},
  {Zamorani}, {Aldcroft}, {Aussel}, {Brunner}, {Capak}, {Cappelluti}, {Civano},
  {Comastri}, {Elvis}, {Feruglio}, {Finoguenov}, {Fruscione}, {Gilli},
  {Hasinger}, {Koekemoer}, {Kartaltepe}, {Ilbert}, {Impey}, {Le Floc'h},
  {Lilly}, {Mainieri}, {Martinez-Sansigre}, {McCracken}, {Menci}, {Merloni},
  {Miyaji}, {Sanders}, {Sargent}, {Schinnerer}, {Scoville}, {Silverman},
  {Smolcic}, {Steffen}, {Santini}, {Taniguchi}, {Thompson}, {Trump}, {Vignali},
  {Urry}, \& {Yan}}]{Fiore09}
{Fiore}, F., {Puccetti}, S., {Brusa}, M., {et~al.} 2009, \apj, 693, 447

\bibitem[{{Freeman} {et~al.}(2002){Freeman}, {Kashyap}, {Rosner}, \&
  {Lamb}}]{Freeman02}
{Freeman}, P.~E., {Kashyap}, V., {Rosner}, R., \& {Lamb}, D.~Q. 2002, \apjs,
  138, 185

\bibitem[{{Frontera} {et~al.}(2007){Frontera}, {Orlandini}, {Landi},
  {Comastri}, {Fiore}, {Setti}, {Amati}, {Costa}, {Masetti}, \&
  {Palazzi}}]{Frontera07}
{Frontera}, F., {Orlandini}, M., {Landi}, R., {et~al.} 2007, \apj, 666, 86

\bibitem[{{Gehrels} {et~al.}(2004){Gehrels}, {Chincarini}, {Giommi}, {Mason},
  {Nousek}, {Wells}, {White}, {Barthelmy}, {Burrows}, {Cominsky}, {Hurley},
  {Marshall}, {M{\'e}sz{\'a}ros}, {Roming}, {Angelini}, {Barbier}, {Belloni},
  {Campana}, {Caraveo}, {Chester}, {Citterio}, {Cline}, {Cropper}, {Cummings},
  {Dean}, {Feigelson}, {Fenimore}, {Frail}, {Fruchter}, {Garmire}, {Gendreau},
  {Ghisellini}, {Greiner}, {Hill}, {Hunsberger}, {Krimm}, {Kulkarni}, {Kumar},
  {Lebrun}, {Lloyd-Ronning}, {Markwardt}, {Mattson}, {Mushotzky}, {Norris},
  {Osborne}, {Paczynski}, {Palmer}, {Park}, {Parsons}, {Paul}, {Rees},
  {Reynolds}, {Rhoads}, {Sasseen}, {Schaefer}, {Short}, {Smale}, {Smith},
  {Stella}, {Tagliaferri}, {Takahashi}, {Tashiro}, {Townsley}, {Tueller},
  {Turner}, {Vietri}, {Voges}, {Ward}, {Willingale}, {Zerbi}, \&
  {Zhang}}]{Gehrels04}
{Gehrels}, N., {Chincarini}, G., {Giommi}, P., {et~al.} 2004, \apj, 611, 1005

\bibitem[{{Giavalisco} {et~al.}(2004){Giavalisco}, {Ferguson}, {Koekemoer},
  {Dickinson}, {Alexander}, {Bauer}, {Bergeron}, {Biagetti}, {Brandt},
  {Casertano}, {Cesarsky}, {Chatzichristou}, {Conselice}, {Cristiani}, {Da
  Costa}, {Dahlen}, {de Mello}, {Eisenhardt}, {Erben}, {Fall}, {Fassnacht},
  {Fosbury}, {Fruchter}, {Gardner}, {Grogin}, {Hook}, {Hornschemeier}, {Idzi},
  {Jogee}, {Kretchmer}, {Laidler}, {Lee}, {Livio}, {Lucas}, {Madau},
  {Mobasher}, {Moustakas}, {Nonino}, {Padovani}, {Papovich}, {Park},
  {Ravindranath}, {Renzini}, {Richardson}, {Riess}, {Rosati}, {Schirmer},
  {Schreier}, {Somerville}, {Spinrad}, {Stern}, {Stiavelli}, {Strolger},
  {Urry}, {Vandame}, {Williams}, \& {Wolf}}]{Giavalisco04}
{Giavalisco}, M., {Ferguson}, H.~C., {Koekemoer}, A.~M., {et~al.} 2004, \apjl,
  600, L93

\bibitem[{{Gilli} {et~al.}(2007){Gilli}, {Comastri}, \& {Hasinger}}]{Gilli07}
{Gilli}, R., {Comastri}, A., \& {Hasinger}, G. 2007, \aap, 463, 79

\bibitem[{{Gilli} {et~al.}(2011){Gilli}, {Su}, {Norman}, {Vignali}, {Comastri},
  {Tozzi}, {Rosati}, {Stiavelli}, {Brandt}, {Xue}, {Luo}, {Castellano},
  {Fontana}, {Fiore}, {Mainieri}, \& {Ptak}}]{Gilli11}
{Gilli}, R., {Su}, J., {Norman}, C., {et~al.} 2011, \apjl, 730, L28

\bibitem[{{Grazian} {et~al.}(2006){Grazian}, {Fontana}, {de Santis}, {Nonino},
  {Salimbeni}, {Giallongo}, {Cristiani}, {Gallozzi}, \& {Vanzella}}]{Grazian06}
{Grazian}, A., {Fontana}, A., {de Santis}, C., {et~al.} 2006, \aap, 449, 951

\bibitem[{{Hall} {et~al.}(2008){Hall}, {Holland}, \& {Turner}}]{Hall08}
{Hall}, D., {Holland}, A., \& {Turner}, M. 2008, in Society of Photo-Optical
  Instrumentation Engineers (SPIE) Conference Series, Vol. 7021, Society of
  Photo-Optical Instrumentation Engineers (SPIE) Conference Series

\bibitem[{{Hasinger}(2008)}]{Hasinger08}
{Hasinger}, G. 2008, \aap, 490, 905

\bibitem[{{Hickox} \& {Markevitch}(2006)}]{Hickox06}
{Hickox}, R.~C. \& {Markevitch}, M. 2006, \apj, 645, 95

\bibitem[{{Hickox} \& {Markevitch}(2007)}]{Hickox07}
{Hickox}, R.~C. \& {Markevitch}, M. 2007, \apjl, 661, L117

\bibitem[{{Hopkins} \& {Beacom}(2006)}]{Hopkins06}
{Hopkins}, A.~M. \& {Beacom}, J.~F. 2006, \apj, 651, 142

\bibitem[{{Kalberla} {et~al.}(2005){Kalberla}, {Burton}, {Hartmann}, {Arnal},
  {Bajaja}, {Morras}, \& {P{\"o}ppel}}]{Kalberla05}
{Kalberla}, P.~M.~W., {Burton}, W.~B., {Hartmann}, D., {et~al.} 2005, \aap,
  440, 775

\bibitem[{{Kuntz} \& {Snowden}(2000)}]{Kuntz00}
{Kuntz}, K.~D. \& {Snowden}, S.~L. 2000, \apj, 543, 195

\bibitem[{{La Franca} {et~al.}(2005){La Franca}, {Fiore}, {Comastri}, {Perola},
  {Sacchi}, {Brusa}, {Cocchia}, {Feruglio}, {Matt}, {Vignali}, {Carangelo},
  {Ciliegi}, {Lamastra}, {Maiolino}, {Mignoli}, {Molendi}, \&
  {Puccetti}}]{Lafranca05}
{La Franca}, F., {Fiore}, F., {Comastri}, A., {et~al.} 2005, \apj, 635, 864

\bibitem[{{Lehmer} {et~al.}(2012){Lehmer}, {Xue}, {Brandt}, {Alexander},
  {Bauer}, {Brusa}, {Comastri}, {Gilli}, {Hornschemeier}, {Luo}, {Paolillo},
  {Ptak}, {Shemmer}, {Schneider}, {Tozzi}, \& {Vignali}}]{Lehmer12}
{Lehmer}, B.~D., {Xue}, Y.~Q., {Brandt}, W.~N., {et~al.} 2012, \apj, 752, 46

\bibitem[{{Luo} {et~al.}(2008){Luo}, {Bauer}, {Brandt}, {Alexander}, {Lehmer},
  {Schneider}, {Brusa}, {Comastri}, {Fabian}, {Finoguenov}, {Gilli},
  {Hasinger}, {Hornschemeier}, {Koekemoer}, {Mainieri}, {Paolillo}, {Rosati},
  {Shemmer}, {Silverman}, {Smail}, {Steffen}, \& {Vignali}}]{Luo08}
{Luo}, B., {Bauer}, F.~E., {Brandt}, W.~N., {et~al.} 2008, \apjs, 179, 19

\bibitem[{{Luo} {et~al.}(2011){Luo}, {Brandt}, {Xue}, {Alexander}, {Brusa},
  {Bauer}, {Comastri}, {Fabian}, {Gilli}, {Lehmer}, {Rafferty}, {Schneider}, \&
  {Vignali}}]{Luo11}
{Luo}, B., {Brandt}, W.~N., {Xue}, Y.~Q., {et~al.} 2011, \apj, 740, 37

\bibitem[{{McQuinn}(2012)}]{Mcquinn12}
{McQuinn}, M. 2012, ArXiv e-prints

\bibitem[{{Mineo} {et~al.}(2012){Mineo}, {Gilfanov}, \& {Sunyaev}}]{Mineo12b}
{Mineo}, S., {Gilfanov}, M., \& {Sunyaev}, R. 2012, ArXiv e-prints

\bibitem[{{Moretti} {et~al.}(2003){Moretti}, {Campana}, {Lazzati}, \&
  {Tagliaferri}}]{Moretti03}
{Moretti}, A., {Campana}, S., {Lazzati}, D., \& {Tagliaferri}, G. 2003, \apj,
  588, 696

\bibitem[{{Moretti} {et~al.}(2005){Moretti}, {Campana}, {Mineo}, {Romano},
  {Abbey}, {Angelini}, {Beardmore}, {Burkert}, {Burrows}, {Capalbi},
  {Chincarini}, {Citterio}, {Cusumano}, {Freyberg}, {Giommi}, {Goad}, {Godet},
  {Hartner}, {Hill}, {Kennea}, {La Parola}, {Mangano}, {Morris}, {Nousek},
  {Osborne}, {Page}, {Pagani}, {Perri}, {Tagliaferri}, {Tamburelli}, \&
  {Wells}}]{Moretti05}
{Moretti}, A., {Campana}, S., {Mineo}, T., {et~al.} 2005, in SPIE Conference
  Series, ed. O.~H.~W. {Siegmund}, Vol. 5898, 360--368

\bibitem[{{Moretti} {et~al.}(2011){Moretti}, {Gastaldello}, {Ettori}, \&
  {Molendi}}]{Moretti11}
{Moretti}, A., {Gastaldello}, F., {Ettori}, S., \& {Molendi}, S. 2011, \aap,
  528, A102

\bibitem[{{Moretti} {et~al.}(2009){Moretti}, {Pagani}, {Cusumano}, {Campana},
  {Perri}, {Abbey}, {Ajello}, {Beardmore}, {Burrows}, {Chincarini}, {Godet},
  {Guidorzi}, {Hill}, {Kennea}, {Nousek}, {Osborne}, \&
  {Tagliaferri}}]{Moretti09}
{Moretti}, A., {Pagani}, C., {Cusumano}, G., {et~al.} 2009, \aap, 493, 501

\bibitem[{{Moretti} {et~al.}(2007){Moretti}, {Perri}, {Capalbi}, {Abbey},
  {Angelini}, {Beardmore}, {Burrows}, {Campana}, {Chincarini}, {Citterio},
  {Cusumano}, {Evans}, {Giommi}, {Goad}, {Godet}, {Guidorzi}, {Grupe}, {Hill},
  {Kennea}, {La Parola}, {Mangano}, {Mineo}, {Morris}, {Nousek}, {Osborne},
  {Page}, {Pagani}, {Racusin}, {Romano}, {Tagliaferri}, \&
  {Tamburelli}}]{Moretti07}
{Moretti}, A., {Perri}, M., {Capalbi}, M., {et~al.} 2007, in SPIE Conference
  Series, ed. S.~L. {O'Dell} \& G.~{Pareschi}, Vol. 6688

\bibitem[{{Mortlock} {et~al.}(2011){Mortlock}, {Warren}, {Venemans}, {Patel},
  {Hewett}, {McMahon}, {Simpson}, {Theuns}, {Gonz{\'a}les-Solares}, {Adamson},
  {Dye}, {Hambly}, {Hirst}, {Irwin}, {Kuiper}, {Lawrence}, \&
  {R{\"o}ttgering}}]{Mortlock11}
{Mortlock}, D.~J., {Warren}, S.~J., {Venemans}, B.~P., {et~al.} 2011, \nat,
  474, 616

\bibitem[{{Revnivtsev} {et~al.}(2005){Revnivtsev}, {Gilfanov}, {Jahoda}, \&
  {Sunyaev}}]{Revnivtsev05}
{Revnivtsev}, M., {Gilfanov}, M., {Jahoda}, K., \& {Sunyaev}, R. 2005, \aap,
  444, 381

\bibitem[{{Salvaterra} {et~al.}(2005){Salvaterra}, {Haardt}, \&
  {Ferrara}}]{Salvaterra05}
{Salvaterra}, R., {Haardt}, F., \& {Ferrara}, A. 2005, \mnras, 362, L50

\bibitem[{{Salvaterra} {et~al.}(2007){Salvaterra}, {Haardt}, \&
  {Volonteri}}]{Salvaterra07}
{Salvaterra}, R., {Haardt}, F., \& {Volonteri}, M. 2007, \mnras, 374, 761

\bibitem[{{Salvaterra} {et~al.}(2012){Salvaterra}, {Haardt}, {Volonteri}, \&
  {Moretti}}]{Salvaterra12}
{Salvaterra}, R., {Haardt}, F., {Volonteri}, M., \& {Moretti}, A. 2012, ArXiv
  e-prints

\bibitem[{{Severgnini} {et~al.}(2012){Severgnini}, {Caccianiga}, \& {Della
  Ceca}}]{Severgnini12}
{Severgnini}, P., {Caccianiga}, A., \& {Della Ceca}, R. 2012, ArXiv e-prints

\bibitem[{{Shull} {et~al.}(2011){Shull}, {Smith}, \& {Danforth}}]{Shull12}
{Shull}, J.~M., {Smith}, B.~D., \& {Danforth}, C.~W. 2011, ArXiv e-prints

\bibitem[{{Snowden} {et~al.}(1998){Snowden}, {Egger}, {Finkbeiner}, {Freyberg},
  \& {Plucinsky}}]{Snowden98}
{Snowden}, S.~L., {Egger}, R., {Finkbeiner}, D.~P., {Freyberg}, M.~J., \&
  {Plucinsky}, P.~P. 1998, \apj, 493, 715

\bibitem[{{Tozzi} {et~al.}(2006){Tozzi}, {Gilli}, {Mainieri}, {Norman},
  {Risaliti}, {Rosati}, {Bergeron}, {Borgani}, {Giacconi}, {Hasinger},
  {Nonino}, {Streblyanska}, {Szokoly}, {Wang}, \& {Zheng}}]{Tozzi06}
{Tozzi}, P., {Gilli}, R., {Mainieri}, V., {et~al.} 2006, \aap, 451, 457

\bibitem[{{Treister} {et~al.}(2006){Treister}, {Urry}, {Van Duyne},
  {Dickinson}, {Chary}, {Alexander}, {Bauer}, {Natarajan}, {Lira}, \&
  {Grogin}}]{Treister06}
{Treister}, E., {Urry}, C.~M., {Van Duyne}, J., {et~al.} 2006, \apj, 640, 603

\bibitem[{{Treister} {et~al.}(2009){Treister}, {Urry}, \&
  {Virani}}]{Treister09}
{Treister}, E., {Urry}, C.~M., \& {Virani}, S. 2009, \apj, 696, 110

\bibitem[{{Ueda} {et~al.}(2003){Ueda}, {Akiyama}, {Ohta}, \& {Miyaji}}]{Ueda03}
{Ueda}, Y., {Akiyama}, M., {Ohta}, K., \& {Miyaji}, T. 2003, \apj, 598, 886

\bibitem[{{Vattakunnel} {et~al.}(2012){Vattakunnel}, {Tozzi}, {Matteucci},
  {Padovani}, {Miller}, {Bonzini}, {Mainieri}, {Paolillo}, {Vincoletto},
  {Brandt}, {Luo}, {Kellermann}, \& {Xue}}]{Vattakunnel12}
{Vattakunnel}, S., {Tozzi}, P., {Matteucci}, F., {et~al.} 2012, \mnras, 420,
  2190

\bibitem[{{Volonteri}(2010)}]{Volonteri10}
{Volonteri}, M. 2010, \aapr, 18, 279

\bibitem[{{Willott} {et~al.}(2010){Willott}, {Delorme}, {Reyl{\'e}}, {Albert},
  {Bergeron}, {Crampton}, {Delfosse}, {Forveille}, {Hutchings}, {McLure},
  {Omont}, \& {Schade}}]{Willott10}
{Willott}, C.~J., {Delorme}, P., {Reyl{\'e}}, C., {et~al.} 2010, \aj, 139, 906

\bibitem[{{Worsley} {et~al.}(2004){Worsley}, {Fabian}, {Barcons}, {Mateos},
  {Hasinger}, \& {Brunner}}]{Worsley04}
{Worsley}, M.~A., {Fabian}, A.~C., {Barcons}, X., {et~al.} 2004, \mnras, 352,
  L28

\bibitem[{{Worsley} {et~al.}(2006){Worsley}, {Fabian}, {Bauer}, {Alexander},
  {Brandt}, \& {Lehmer}}]{Worsley06}
{Worsley}, M.~A., {Fabian}, A.~C., {Bauer}, F.~E., {et~al.} 2006, \mnras, 368,
  1735

\bibitem[{{Xue} {et~al.}(2011){Xue}, {Luo}, {Brandt}, {Bauer}, {Lehmer},
  {Broos}, {Schneider}, {Alexander}, {Brusa}, {Comastri}, {Fabian}, {Gilli},
  {Hasinger}, {Hornschemeier}, {Koekemoer}, {Liu}, {Mainieri}, {Paolillo},
  {Rafferty}, {Rosati}, {Shemmer}, {Silverman}, {Smail}, {Tozzi}, \&
  {Vignali}}]{Xue11}
{Xue}, Y.~Q., {Luo}, B., {Brandt}, W.~N., {et~al.} 2011, \apjs, 195, 10

\bibitem[{{Xue} {et~al.}(2012){Xue}, {Wang}, {Brandt}, {Luo}, {Alexander},
  {Bauer}, {Comastri}, {Fabian}, {Gilli}, {Lehmer}, {Schneider}, {Vignali}, \&
  {Young}}]{Xue12}
{Xue}, Y.~Q., {Wang}, S.~X., {Brandt}, W.~N., {et~al.} 2012, ArXiv e-prints

\end{thebibliography}

\end{document}